\documentstyle[prb,aps]{revtex}
\topmargin - 1 cm

\begin{document}

\title{Manifestation of exciton Bose condensation in induced
two-phonon emission and Raman scattering}

\author{Yu.E.Lozovik\cite{E} and A.V.Poushnov}

\address{Institute of Spectroscopy, Russian  Academy  of  Sciences,
142092, Troitsk, Moscow region, Russia}

\maketitle

\begin{abstract}
The unusual two-photon emission by Bose-condensed excitons caused by
{\it simultaneous} recombination of {\it two} excitons with {\it opposite}
momenta leaving the occupation numbers of excitonic states with momenta
${\bf p}\neq 0$ unchanged (below coherent two-exciton recombination) is
investigated. Raman scattering accompanied by the analogous
two-exciton recombination (or creation) is also analyzed. The excess
momentum equal to the change of the electromagnetic field momentum in these
processes can be transferred to phonons or impurities.  The processes under
consideration take place if there is Bose condensation in the interacting
exciton system, and, therefore, can be used as a new method to reveal
exciton Bose condensation.  If the frequency of the incident
light $\omega<2\Omega$ ($\Omega$ is the frequency corresponding to the
recombination of an exciton with {\bf p}=0), the coherent two-exciton
recombination with the excess momentum elastically transferred to impurities
leads to the appearance of the spectral line $2\Omega-\omega$ corresponding
to the induced two-photon emission. In this case the anti-Stokes line on
frequency $\omega+2\Omega$ also appears in the Raman spectrum.
If $\omega>2\Omega$, there are both Stokes and anti-Stokes lines on
frequencies $\omega\pm2\Omega$ in the Raman spectrum. The induced
two-photon emission is impossible in this case. The spectral lines mentioned
above have phonon replicas on frequencies
$|\omega\pm (2\Omega-n\omega^s_0)|$ corresponding to the
transmission of the excess momentum (partially or as a whole) to optical
phonons of frequency $\omega^s_0$ ($n$ is an integer number).  The
quantitative estimation shows that the light corresponding to the coherent
two-exciton recombination can be experimentally observed in $Cu_2O$.
\end{abstract}

PACS: 71.35.-y; 71.35.Lk; 78.20.Wc; 78.60.Ya

\newpage

\section{Introduction}

\ \ \ One of the most interesting collective properties of excitons is their
possible Bose condensation and superfluidity.\cite{BLATT}$^{-}$\cite{BOSE}
There are interesting
reports about the experimental observation of Bose condensation and
superfluidity of excitons in $Cu_2O$ based on the detected changes in the
exciton luminescence spectrum \cite{TWO}$^,$\cite{PARA} and the ballistic
exciton transport \cite{TRANSPORT}$^,$\cite{BENSON} (see also the discussion
in Refs.\cite{BENSON}$^-$\cite{TIKHODEEV}). There are also the reports about
the observation of condensation of indirect excitons in coupled quantum well
structures in strong magnetic field \cite{BUTOV94} (see also theoretical
investigations of superfluidity in coupled quantum wells in
Refs.\cite{YUDSON}--\cite{Klyuchnik}; strong magnetic field effects are
considered in Ref.\cite{LERNER}). In view of the strong interest to the
search for exciton Bose condensation the new methods of the detection of Bose
condensation of excitons seems to be a very important problem.

If excitons are Bose-condensed, the average of the destruction (or creation)
operator of an exciton with momentum ${\bf p}=0$ is not zero:

\begin{eqnarray}
\label{Q}
\langle N-1|Q_0|N \rangle=\langle N+1|Q^+_0|N \rangle=\sqrt{N_0}.
\end{eqnarray}
Here $|N \rangle$ is the ground state of the exciton system corresponding
to average number of excitons $N$, $Q_0$ is the destruction operator of
an exciton with ${\bf p}=0$, $N_0$ is the number of condensate excitons.

From averages (\ref{Q}) it is easy to see that after recombination
or creation of an exciton with ${\bf p}=0$ the Bose-condensed exciton
system, being in the ground state, also appears in the ground state which
differs from the initial one only by the average number of excitons
with ${\bf p}=0$. The
condensate peak in the exciton luminescence spectrum on frequency
$\Omega=[E_0(N)-E_0(N-1)]/\hbar$ corresponds to the recombination
of excitons with ${\bf p}=0$ (here $E_0(N)$ is the energy of the ground
state of the exciton system).

If exciton-exciton interaction is present, not only averages
(\ref{Q}) but also the product of two destruction (or creation) operators
of excitons with opposite momenta averaged on the ground state of the
exciton system is not zero:

\begin{eqnarray}
\label{QQ}
\langle N-2|Q_{-p}Q_{p}|N \rangle\neq 0;
\nonumber \\
\langle N+2|Q^+_{-p}Q^+_{p}|N \rangle\neq 0.
\end{eqnarray}

The unusual optical properties of excitons in Bose-condensed state due
to nonzero averages (\ref{QQ}) are investigated in this paper. It is
shown that the interaction of Bose-condensed excitons with
electromagnetic field leads to the possibility of the simultaneous
recombination (or creation) of two excitons with opposite momenta
corresponding to averages (\ref{QQ}). These processes leave the
occupation numbers of excitons with ${\bf p}\neq 0$ unchanged. The only
difference between initial and final states of the exciton system is the
average number of excitons with ${\bf p}=0$.
Below the recombination (creation) of two excitons with opposite momenta
corresponding to averages (\ref{QQ}) will be called {\it coherent two-exciton
recombination (creation)}.

The coherent two-exciton recombination can accompany the induced
two-photon emission and Raman scattering. Raman scattering can
also be accompanied by the coherent two-exciton creation. In these processes
the momentum of the exciton-photon system is not conserved. In fact, the
momentum of the exciton system remains unchanged, while the momentum of
the electromagnetic field is changed. The excess momentum equal to the change
of the electromagnetic field momentum in these processes is transferred to
phonons or impurities.\cite{1}

If the frequency of the incident light $\omega<2\Omega$, the spectral line
on frequency $2\Omega-\omega$ corresponding to the induced impurity-assisted
two-photon emission and anti-Stokes line on frequency $\omega+2\Omega$
corresponding to impurity-assisted Raman scattering appear.
Both lines correspond to the coherent two-exciton recombination:
the energy of the initial state of exciton system  exceeds the energy
of the final state by $2\hbar\Omega$, where $\Omega$ is the frequency
corresponding to the recombination of an exciton with ${\bf p}=0$.
The transmission of the excess momentum to impurities is supposed to be
elastic. If $\omega>2\Omega$, there are anti-Stokes line $\omega+2\Omega$
corresponding to the coherent two-exciton recombination and Stokes line
$\omega-2\Omega$ corresponding to the coherent two-exciton creation
in the Raman spectrum. The induced two-photon emission is
not possible in this case. The appearance of the lines on frequencies
$|\omega\pm 2\Omega|$ is possible only if excitons are Bose-condensed. If
exciton system is in the normal state, these lines are absent.

Analogous effects take place if the excess momentum is taken by phonons
(partially or as a whole). If phonons are optical, there are phonon replicas
of spectral lines $|\omega\pm 2\Omega|$ on frequencies
$|\omega\pm(2\Omega-n\omega^s_0)|$ ($\omega^s_0$ is the frequency of
the optical phonons, $n$ is an integer number).

The quantitative estimation for excitons in $Cu_2O$ shows that the
experimental observation of the spectral line on frequency
$2(\Omega-\omega^s_0)-\omega$ corresponding to the coherent two-exciton
recombination is possible and, therefore, can be used to detect exciton Bose
condensation.

\section{The impurity-assisted two-photon emission}
\label{IMP}

\ \ \ Effective Hamiltonian $\hat H_X$ responsible for the optical
impurity-assisted exciton recombination can be represented in the following
form (see Appendix \ref{D}):

\begin{eqnarray}
\label{X}
\hat H_X=\hat X+\hat X',
\nonumber \\
\hat X=\sum_{jpq}^{}\left(X^j_{pq}Q_pc^+_{q}+h.c.\right),
\nonumber \\
\hat X'=\sum_{jpq}^{}\left(X'^j_{pq}Q_pc_{q}+h.c.\right),
\end{eqnarray}
where $X^j_{pq}=i\sqrt{2\pi \omega_q}({\bf e^*d}^j_{pq})$,
$X'^j_{pq}=-i\sqrt{2\pi \omega_q}({\bf ed'}^j_{pq})$. Here
$Q_p$ is the destruction operator of an exciton with momentum ${\bf p}$,
$c_{q}$ is the destruction operator of a photon with momentum ${\bf q}$
($\omega_q$ and ${\bf e}$ are the frequency and the polarization vector
of the photon). Matrix elements
${\bf d}^j_{pq}({\bf d'}^j_{pq})$ are responsible for
the optical recombination of an exciton with momentum ${\bf p}$ accompanied
by the transmission of excess momentum ${\bf p}\mp{\bf q}$ to impurity $j$
(see Appendix \ref{D}). Of course, in the first order on $\hat X'$ the exciton
recombination is impossible due to the energy conservation.

Let us consider Bose-condensed excitons at $T=0$, i.e. in vacuum state
$|i\rangle_{exc}=|0\rangle_{exc}$ with respect to quasiparticles (the case
$T\neq 0$ will be considered in another paper).
Because the excitons are in the coherent state, the
recombination (creation) of two excitons with equal and opposite momenta
accompanied by the excitation of no quasiparticles is possible
(see averages (\ref{QQ})). Therefore, two-photon processes (two-photon
emission and Raman scattering) accompanied by the coherent two-exciton
recombination (creation) can take place. These processes correspond to the
transition of the exciton system to the final state
$|f\rangle_{exc}=|0\rangle_{exc}$ which differs from $|i\rangle_{exc}$ only
by the average number of excitons with ${\bf p}=0$.

In this section we shall consider the induced two-photon emission accompanied
by the coherent two-exciton recombination with the excess
momentum transferred to impurities. The case of the transmission of the excess
momentum to phonons is considered in Sec.\ref{PHN}.
Raman scattering with coherent two-exciton recombination (creation) is
discussed in Sec.\ref{RAMAN}.

If the scattering of excitons on impurities is elastic, the energy
conservation law for the induced two-photon emission accompanied by the
coherent two-exciton recombination is

\begin{eqnarray}
\label{energy}
2\Omega=\omega+\omega',
\end{eqnarray}
where $\omega$ and $\omega'$ are the energies of the emitted photons (here
and further $\hbar=1$ if it is not specially mentioned), $\Omega$ is the
energy corresponding to the recombination of an exciton with ${\bf p}=0$.
In case of the {\it induced} two-photon emission one of the frequencies
($\omega$) in Eq.(\ref{energy}) coincides with the frequency of the incident
light. Thus, the spectral line on frequency $2\Omega-\omega$ corresponding
to the induced two-photon emission by excitons appears at the temperature
below the Bose-Einstein condensation point.

The matrix element for the two-photon emission under consideration is

\begin{eqnarray}
(\hat H_X)_{fi}=\sum_{\nu}^{}\left(
\frac{(\hat X)_{f\nu}(\hat X)_{\nu i}}{\omega_{i\nu}-\omega+i\delta}+
\frac{(\hat X)_{f\nu}(\hat X)_{\nu i}}{\omega_{i\nu}-\omega'+i\delta}
\right)
\end{eqnarray}
where $|i\rangle=|0\rangle_{exc}|0\rangle_{phot}$ and
$|f\rangle=|0\rangle_{exc}|1_k,1_{k'}\rangle_{phot}$ are initial and
final states of the exciton-photon system, ${\bf k}$ and ${\bf k'}$ are
the momenta of the emitted photons, $\omega_{i\nu}$ is the energy difference
between initial and intermediate states of the exciton system.
Frequencies $\omega$ and $\omega'$ are related by energy conservation
law (\ref{energy}).

The two-photon emission under consideration takes place through the
intermediate states of two types with respect to the exciton occupation
numbers:

{\bf I}. After the recombination of an  exciton with momentum
${\bf p}=0$ the exciton system transfers into the intermediate state
containing no quasiparticles. This intermediate state differs from the
initial one only by the average number of excitons with ${\bf p}=0$. During
the following transition into the final state  other exciton with
${\bf p}=0$ recombines.

{\bf II}. After the recombination of an exciton with momentum
${\bf p}\neq 0$ the exciton system transfers into the {\it excited}
intermediate state containing a quasiparticle with momentum
${\bf -p}$. During the following transition of the exciton system into the
final state an  exciton with the opposite momentum recombines and the
quasiparticle  disappears. Thus, there are no quasiparticles in the final
state of the exciton system.

The amplitudes of the intermediate states of the exciton system and the
matrix elements for the corresponding transitions are

\begin{eqnarray}
\label{I,II}
&&|\nu\rangle_{exc}=[\delta_p+(1-\delta_p)O^+_{-p}]|0\rangle_{exc},
\nonumber \\ &&
(\hat X)_{\nu i}=X^i_{pk}[\sqrt{N_0}\delta_p+(1-\delta_p)v_p],
\nonumber \\ &&
(\hat X)_{f\nu}=X^j_{-pk'}[\sqrt{N_0}\delta_p+(1-\delta_p)u_p],
\end{eqnarray}
where $\delta_p=1$ at $p=0$; $\delta_p=0$ at $p\neq 0$. Here $O^+_p$
is the creation operator of a quasiparticle in the exciton system defined by
the Bogoliubov transformation (we consider the dilute exciton system):

\begin{eqnarray}
\label{uv}
O^+_p=u_pQ^+_{p}-v_pQ_{-p}.
\end{eqnarray}

In process of the two-photon emission the change of the momentum of the
electromagnetic field is ${\bf k'}+{\bf k}$, where ${\bf k}$ and ${\bf k'}$
are the momenta of the emitted photons. Since the momentum of the exciton
system remains unchanged in this process, the excess momentum
$\delta{\bf k}={\bf -(k+k')}$ is transferred to impurities.

At first we shall consider the case when two {\it different} impurities take
the excess momentum corresponding to the induced two-photon emission
accompanied by the coherent two-exciton recombination. Let impurities
$i$ and $j$ take momenta ${\bf p-k}$ and ${\bf -(k'+p)}$ correspondingly.
There can be only quasiparticles with momenta $\pm {\bf p}$ and
$\pm({\bf p-k+k'})$ in the exciton system in this case. Diagrams
corresponding to such two-photon emission (see Appendix \ref{A}) are shown in
Fig.1.

The matrix element for the two-photon emission with the excess momentum
transferred to impurities $i$ and $j$ has the following form

\begin{eqnarray}
\label{Mij}
&&(\hat H^{ij}_X)_{fi}=
\left[\frac{X^j_{-pk'}X^i_{pk}}{\Omega-\epsilon_p-\omega+i\Gamma_p/2}+
\frac{X^i_{pk}X^j_{-pk'}}{\Omega-\epsilon_p-\omega'+i\Gamma_p/2}\right]\times
\nonumber \\ && \times
\left(N_0\delta_p+(1-\delta_p)u_pv_p\right)+
\nonumber \\ && +
\left[\frac{X^i_{p-q,k'}X^j_{-p+q,k})}{\Omega-\epsilon_{p-q}-
\omega+i\Gamma_{p-q}/2}+
\frac{X^j_{-p+q,k}X^i_{p-q,k'}}{\Omega-\epsilon_{p-q}-
\omega'+i\Gamma_{p-q}/2}\right]\times
\nonumber \\ && \times
\left(N_0\delta_{p-q}+(1-\delta_{p-q})u_{p-q}v_{p-q}\right),
\end{eqnarray}
where $\Gamma_p$ is the width of the energy level of the exciton system
corresponding to the quasiparticle with momentum $p$; ${\bf q=k-k'}$.
In Eq.(\ref{Mij}) we have taken into account that the difference
between the energies of the exciton system in initial and
intermediate states is $\omega_{i\nu}=\Omega-\epsilon_p$. At $p\neq 0$
the quantity $\epsilon_p$ is the energy of the quasiparticle in the
intermediate state of the exciton system; $\epsilon_p=0$ at $p=0$.

Using energy conservation law (\ref{energy}), matrix element (\ref{Mij})
can be expressed in terms of the anomalous Green function of Bose-condensed
excitons:

\begin{eqnarray}
\label{ij}
&&(\hat H^{ij}_X)_{fi}=-\left[2\pi iN_0(\Omega-\omega)\delta_p-
(1-\delta_p)\hat G_p(\Omega-\omega)\right]X^j_{-pk'}X^i_{pk}-
\nonumber \\ && -
\left[2\pi iN_0(\Omega-\omega)\delta_{p-q}-
(1-\delta_{p-q})\hat G_{p-q}(\Omega-\omega)\right]X^i_{p-q,k'}X^j_{-p+q,k},
\end{eqnarray}
where

\begin{eqnarray}
\label{G}
\hat G_p(\omega)=\frac{u_pv_p}{\omega-\epsilon_p+i\Gamma_p/2}-
\frac{u_pv_p}{\omega+\epsilon_p-i\Gamma_p/2}
\end{eqnarray}
is the anomalous Green function of the dilute Bose-condensed exciton
system at $T=0$ (see Appendix \ref{A}),

\begin{eqnarray}
\label{N0}
N_0(\omega)=\frac{N_0}{\pi}\frac{\Gamma_0/2}{\omega^2+\Gamma^2_0/4}.
\end{eqnarray}

For the differential cross-section of the induced two-photon emission
with the excess momentum transferred to two different impurities
one has

\begin{eqnarray}
\label{dsij}
d\sigma^X_2=\frac{\omega(2\Omega-\omega)^3}{c^4}
\sum_{(ij)}^{}\sum_{p}|(s^{ij}_p)_{nm}e'^*_ne^*_m|^2do',
\end{eqnarray}
where

\begin{eqnarray}
\label{tij}
&&(s^{ij}_p)_{nm}=\left[2\pi iN_0(\Omega-\omega)\delta_p-
(1-\delta_p)\hat G_p(\Omega-\omega)\right](d^j_{-pk'})_n(d^i_{pk})_m+
\nonumber \\ && +
\left[2\pi iN_0(\Omega-\omega)\delta_{p-q}-
(1-\delta_{p-q})\hat G_{p-q}(\Omega-\omega)\right]
(d^i_{p-q,k'})_n(d^j_{-p+q,k})_m.
\end{eqnarray}

At $|\Omega-\omega|\gg\Gamma_0$ the terms $\sim N_0$ give a negligibly
small contribution to the cross-section (\ref{dsij}). The cross-section is
proportional to the chemical potential of excitons in this case,
because $\hat G_p(\omega)\sim\mu$. Therefore, if
$|\Omega-\omega|\gg\Gamma_0$, cross-section (\ref{dsij}) does not directly
depend on factor $N_0$ of macroscopic filling of the state with ${\bf p}=0$.
It is defined by anomalous averages (\ref{QQ}) caused by the existence of
Bose-condensate in the system of {\it interacting} excitons. The chemical
potential of three-dimentional {\it ideal} Bose gas is zero below
Bose-Einstein condensation point. Thus, at $|\Omega-\omega|\gg\Gamma_0$ the
{\it induced} two-photon emission accompanied by the coherent two-exciton
recombination with the excess momentum transferred to two different
impurities is possible only in the {\it nonideal} Bose gas of excitons.

For weakly interacting Bose gas of excitons at $T=0$ the anomalous Green
function is
$\hat G_p(\omega)=-\mu/[\omega^2-(\epsilon_p-i\Gamma_p/2)^2]$, where
$\epsilon_p=\sqrt{\xi_p^2-\mu^2}$. Here $\xi_p=p^2/(2m)+\mu$; $\mu$ is the
chemical potential of the excitons and $m$ is the exciton mass.\cite{GIMP}
If the condition $|\Omega-\omega|=\epsilon_p$ is fulfilled at $p\gg q$, we
can suppose that $\epsilon_{p-q}\simeq\epsilon_p$ in tensor (\ref{tij}).

Replacing the summation on ${\bf p}$ by ${\bf p}$-integration
in Eq.(\ref{dsij}) yields

\begin{eqnarray}
\label{i+i}
&&\sum_{p}|(s^{ij}_p)_{nm}e'^*_ne^*_m|^2=
\mu^2\int\frac{d^3p}{(2\pi)^3}
\biggl\vert\frac{({\bf e'^*d}^j_{-pk'})({\bf e^*d}^i_{pk})+
({\bf e'^*d}^i_{pk'})({\bf e^*d}^j_{-pk})}{
(\Omega-\omega-i\Gamma_p/2)^2-\epsilon_p^2}\biggl\vert^2.
\end{eqnarray}

Integral (\ref{i+i}) diverges if $\Gamma_p \rightarrow 0$. Thus, if
$|\Omega-\omega|\gg\Gamma_p$, the main contribution to the cross-section
of the induced two-photon emission is given by the resonant levels
corresponding to quasiparticles with energies $\epsilon_p\sim|\Omega-\omega|$
in the Bose-condensed exciton system.\cite{GP}
In this case matrix elements ${\bf d}^j_{-pk'}$; ${\bf d}^i_{pk}$ can be
replaced by their values at momentum ${\bf p}_X$ defined by the
condition $\epsilon(p_X)=|\Omega-\omega|$. Replacing the {\bf p}-integration
by the integration on $t=\xi_p/\mu$ yields

\begin{eqnarray}
\label{19}
\sum_{p}^{}|(s^{ij}_p)_{nm}e'^*_ne^*_m|^2=
\frac{2^{1/2}m^{3/2}\mu^{-1/2}}{2\pi^2}
|[d^j_n(\omega'_X)d^i_m(\omega_X)+d^i_n(\omega'_X)d^j_m(\omega_X)]
e'^*_ne^*_m|^2
\times \nonumber \\ \times
\int_{1}^{\infty}
\frac{dt\sqrt{t-1}}{(\alpha_{X+}^2+1-t^2)(\alpha_{X-}^2+1-t^2)},
\end{eqnarray}
where $\alpha_{X\pm}=\alpha_X \pm i\gamma_p$, $\alpha_X=|\Omega-\omega|/\mu$,
$\gamma_p=\Gamma_p/(2\mu)$. Here \\
${\bf d}^i(\omega_X)=\int {\bf d}^i(p_X,k)do_{p_X}/(4\pi)$ and
${\bf d}^i(\omega'_X)=\int {\bf d}^i(p_X,k')do_{p_X}/(4\pi)$
are the matrix elements averaged over ${\bf p}_X$ directions.

Integral in Eq.(\ref{19}) can be represented as the sum of two
integrals each of which does not diverge at $\gamma_p\rightarrow 0$:

\begin{eqnarray}
\label{beta}
\int_{1}^{\infty}
\frac{dt\sqrt{t-1}}{(\alpha_{X+}^2+1-t^2)
(\alpha_{X-}^2+1-t^2)}=
\frac{1}{\beta_+^2-\beta_-^2}\left[
\int_{1}^{\infty}\frac{dt\sqrt{t-1}}{t^2-\beta_+^2}-
\int_{1}^{\infty}\frac{dt\sqrt{t-1}}{t^2-\beta_-^2}
\right],
\end{eqnarray}
where $\beta_\pm^2=\alpha_{X\pm}^2+1$. Thus, the integration in the right side
of Eq.(\ref{beta}) can be done supposing
$\beta^2_\pm=\beta^2\pm i\delta$, $\delta=0+$. One has

\begin{eqnarray}
\sum_{p}^{}|(s^{ij}_p)_{nm}e'^*_ne^*_m|^2=
\frac{2^{1/2}m^{3/2}\mu^{-1/2}}{8\pi\alpha_X\gamma_X}
\frac{(\sqrt{\alpha^2_X+1}-1)^{1/2}}{\sqrt{\alpha^2_X+1}} \times
\nonumber \\ \times
|[d^j_n(\omega'_X)d^i_m(\omega_X)+d^i_n(\omega'_X)d^j_m(\omega_X)]
e'^*_ne^*_m|^2,
\end{eqnarray}
where $\gamma_X=\Gamma_X/(2\mu)$. Here $\Gamma_X$ is the inverse lifetime
of a quasiparticle with energy $\epsilon(p_X)=|\Omega-\omega|$ in
the Bose-condensed exciton system.

Inserting the obtained expression in Eq.(\ref{dsij}) the summation over
impurities can be done. Supposing all the impurities to be identical yields

\begin{eqnarray}
&&d\sigma^X_2=
\nonumber \\ &&
\frac{\omega(2\Omega-\omega)^3}{4\pi c^4}\frac{2^{1/2}m^{3/2}\mu^{-1/2}
(\sqrt{\alpha^2_X+1}-1)^{1/2}}{\alpha_X\gamma_X\sqrt{\alpha_X^2+1}}
N(N-1)|d_n(\omega'_X)d_m(\omega_X)e'^*_ne^*_m|^2do',
\nonumber \\
\end{eqnarray}
where $N$ is the number of the impurities per unit volume.

If the excitons and the impurities are in the isotropic medium and the
exciton scattering on the impurities is also isotropic, one has
$d^2_n(\omega_X)={\bf d}^2(\omega_X)/3$. If the incident light is polarized
along some direction, the equality
$|e^*_md_m(\omega_X)|^2={\bf d}^2(\omega_X)/3$ is valid,
because in case of the {\it induced} two-photon emission photon
$\omega$ is identical to the incident one. Summing over the polarization of
photon $\omega'$ and integrating over the direction of its momentum
yields the total cross-section of the induced two-photon emission with
the coherent two-exciton recombination and the excess momentum transferred to
two different impurities

\begin{eqnarray}
\label{sX2}
\sigma^X_2=
\frac{\omega(2\Omega-\omega)^32^{5/2}m^{3/2}\mu^{1/2}}{9 c^4\alpha_X\Gamma_X}
\frac{(\sqrt{\alpha_X^2+1}-1)^{1/2}}{\sqrt{\alpha_X^2+1}}N(N-1)
{\bf d}^2(\omega_X){\bf d}^2(\omega'_X).
\end{eqnarray}

Now we shall consider the induced two-photon emission with the coherent
two-exciton recombination accompanied by the transmission of excess
momentum $\delta{\bf k}$ to {\it one} of impurities (for example, to
impurity $j$) as a whole (see also Ref.\cite{POUSHNOV}). In this process the
states of the other impurities remain unchanged, that is why the momentum
of a quasiparticle in the intermediate state of the exciton system
{\it does not depend} on the final state of impurity $j$. Therefore, the
matrix element for such two-photon emission is given by the expression

\begin{eqnarray}
\label{1}
(\hat H^{jj}_X)_{fi}=\sum_{p}^{}
\left(\frac{X^j_{-pk'}X^j_{pk}}{\Omega-\epsilon_p-\omega+i\Gamma_p/2}+
\frac{X^j_{-pk}X^j_{pk'}}{\Omega-\epsilon_p-\omega'+i\Gamma_p/2}\right)
\times \nonumber \\ \times
[N_0\delta_p+(1-\delta_p)u_pv_p],
\end{eqnarray}
in which the summation is performed over {\it all possible} momenta ${\bf p}$
of the quasiparticle in the intermediate state of the exciton system.
Diagrams corresponding to the two-photon emission under consideration
are shown in Fig.2. The sum of the diagrams with all possible momenta
${\bf p}$ corresponds to matrix element (\ref{1}) (see Appendix \ref{A}).

For the differential cross-section of the induced two-photon emission with
coherent two-exciton recombination and the excess momentum transferred
to one of the impurities as a whole one has

\begin{eqnarray}
\label{f1i}
d\sigma^X_1=\frac{\omega(2\Omega-\omega)^3}{c^4}
\sum_{j}^{}|(s^{jj})_{nm}e'^*_ne^*_m|^2do',
\end{eqnarray}
where tensor $(s^{jj})_{nm}$ is given by the relation

\begin{eqnarray}
\label{tensor}
(s^{jj})_{nm}=\sum_{p}^{}\left[2\pi iN_0(\Omega-\omega)\delta_p-
(1-\delta_p)\hat G_p(\Omega-\omega)\right](d^j_{-pk'})_n(d^j_{pk})_m.
\end{eqnarray}
Here $\hat G_p(\omega)$ is the anomalous Green function of Bose-condensed
excitons (see Eq.(\ref{G})), $N_0(\omega)$ is defined by Eq.(\ref{N0}).

If the frequency of the incident light satisfies the condition
$|\Omega-\omega|\gg\Gamma_0$, the terms $\sim N_0$ give a negligibly small
contribution in tensor (\ref{tensor}). Cross-section (\ref{f1i}) as well as
cross-section (\ref{dsij}) (see above) does not directly depend on factor
$N_0$ of macroscopic filling of the exciton state with ${\bf p}=0$ in this
case. Therefore, the induced two-photon emission by the {\it ideal} Bose gas
of excitons with the coherent two-exciton recombination and transmission of
the excess momentum to one of the impurities as a whole is impossible if
$|\Omega-\omega|\gg\Gamma_0$.

Replacing the summation on ${\bf p}$ by {\bf p}-integration yields

\begin{eqnarray}
\label{tens}
(s^{jj})_{nm}=\mu\int_{0}^{\infty}\frac{d^3p}{(2\pi)^3}
\frac{(d^j_{-pk'})_n(d^j_{pk})_m}{
(\Omega-\omega+i\Gamma_p/2-\epsilon)(\Omega-\omega-i\Gamma_p/2+\epsilon_p)}.
\end{eqnarray}

If $|\Omega-\omega|\gg\Gamma_p$, the main contribution to the tensor of the
two-photon emission with transmission of the excess momentum to one of the
impurities as a whole is given by the intermediate states corresponding
to quasiparticles with energies $\epsilon_p\sim |\Omega-\omega|$.
In this case ${\bf d}^j_{pk}$; ${\bf d}^j_{-pk'}$ can be replaced by
${\bf d}^j(\omega_X)$ and ${\bf d}^j(\omega'_X)$. The definition of
${\bf d}^j(\omega_X)$ and ${\bf d}^j(\omega'_X)$ is after Eq.(\ref{19}). At
the same time,
integral (\ref{tens}) does not diverge even at $\Gamma_p\rightarrow 0$,
contrary to the case when two different impurities take the excess momentum.
Thus, $\Gamma_p$ can be replaced by $\delta=0+$ if
$|\Omega-\omega|\gg\Gamma_p$. Replacing the {\bf p}-integration by the
integration on $\xi_p$ we have:

\begin{eqnarray}
\label{d}
&&(s^{jj})_{nm}=-d^j_n(\omega'_X)d^j_m(\omega_X)
\frac{2^{1/2}m^{3/2}\mu}{2\pi^2}
\int_{\mu}^{\infty}
\frac{d\xi_p\sqrt{\xi_p-\mu}}{(\Omega-\omega)^2-\epsilon_p^2+i\delta}.
\end{eqnarray}

After the substitutions $t=\xi_p/\mu$ and $u=t-1$ the integration can be
easily done. The tensor of the two-photon emission with the coherent
two-exciton recombination and the transmission of the excess momentum to
impurity $j$ is

\begin{eqnarray}
\label{t}
(s^{jj})_{nm}=\frac{2^{1/2}m^{3/2}\mu^{1/2}}{4\pi\sqrt{\alpha^2_X+1}}
\left[(\sqrt{\alpha^2_X+1}+1)^{1/2}+i(\sqrt{\alpha^2_X+1}-1)^{1/2} \right]
d^j_n(\omega'_X)d^j_m(\omega_X).
\end{eqnarray}
where $\alpha_X=|\Omega-\omega|/\mu$.

Inserting tensor (\ref{t}) in Eq.(\ref{f1i}) the
summation on the impurities can be done. Integrating on the polarization
and the direction of photon $\omega'$ yields the overall cross-section
of the induced two-photon emission with transmission of the excess momentum
to one of the impurities as a whole

\begin{eqnarray}
\label{sX1}
\sigma^X_1=\frac{\omega(2\Omega-\omega)^3}{c^4}
\frac{2m^3\mu N}{
9\pi\sqrt{\alpha^2_X+1}}{\bf d}^2(\omega_X){\bf d}^2(\omega'_X).
\end{eqnarray}

Therefore, the cross-section of the induced impurity-assisted two-photon
emission accompanied by the coherent two-exciton recombination $\sigma^X$
is the sum of two terms:

\begin{eqnarray}
\sigma^X(\omega)=\sigma^X_1(\omega)+\sigma^X_2(\omega),
\end{eqnarray}
where $\sigma^X_1(\omega)$ and $\sigma^X_2(\omega)$ are the cross-sections
of the induced two-photon scattering with the excess momentum transferred to
one of the impurities as a whole and to two different impurities
correspondingly (see Eqs.(\ref{sX1}),(\ref{sX2})). Despite
$\sigma^X_1/\sigma^X_2\sim 1/N$; $N\gg 1$, the ratio between these
cross-sections can not be estimated in general case, because it essentially
depends on other parameters involved in the problem:

\begin{eqnarray}
\label{1/2}
\frac{\sigma^X_1}{\sigma^X_2}=
\frac{m^{3/2}\mu^{1/2}}{2^{3/2}\pi\hbar^2N\tau^X}
\frac{\alpha_X}{\left(\sqrt{\alpha_X^2+1}-1\right)^{1/2}},
\end{eqnarray}
where $\tau^X$ is the lifetime of quasiparticle with energy
$\epsilon(p_X)=|\Omega-\omega|$ in Bose-condensed exciton system.
Eq.(\ref{1/2}) is written in ordinary units.

\section{Phonon replicas}
\label{PHN}

\ \ \ The excess momentum corresponding to the two-photon emission
accompanied by the coherent two-exciton recombination can be transferred not
only to impurities but also to phonons. Thus, two more types of two-photon
emission accompanied by the coherent two-exciton recombination are possible:
1) The excess momentum is transferred to phonons as a whole; 2) An
impurity takes some part of the excess  momentum. The rest of it is
transferred to a phonon. It is easy to see that in both cases the momentum of
the quasiparticle in the intermediate state of the exciton system depends on
the final state of the phonon subsystem of the crystal. Therefore, the
two-photon emission with the excess momentum transferred to phonons
(or to both a phonon and an impurity) is analogous to that accompanied
by the transmission of the excess momentum to two different impurities.

In general, the optical recombination of an exciton can be assisted by an
arbitrary number of phonons. We shall restrict ourselves to the
consideration of an exciton assisted by one phonon. In this case the
effective Hamiltonian responsible for the optical recombination (creation) of
excitons assisted by a phonon can be represented in the form analogous
to that of Hamiltonian (\ref{X}) in Sec.\ref{IMP} (see Appendix \ref{D}):

\begin{eqnarray}
\label{HL}
&&
\hat H_L=\hat L+\hat L',
\nonumber \\ &&
\hat L=\sum_{pq}^{}
\left[L^>_{pq}Q_pc^+_{q}b^+_{p-q}+L^<_{pq}Q_pc^+_{q}b_{-p+q}+h.c.\right],
\nonumber  \\ &&
\hat L'=\sum_{pq}^{}
\left[L'^>_{pq}Q_pc_{q}b^+_{p+q}+L'^<_{pq}Q_pc_{q}b_{-p-q}+h.c.\right],
\end{eqnarray}
where $L^{>(<)}_{pq}=i\sqrt{2\pi \omega_q}({\bf e^*f}^{>(<)}_{pq})$,
$L'^{>(<)}_{pq}=-i\sqrt{2\pi \omega_q}({\bf ef}'^{>(<)}_{pq})$. Here $b_p$
is the destruction operator of a phonon. Effective matrix elements
${\bf f}_{pq}^{>(<)}$ and ${\bf f}'^{>(<)}_{pq}$ are responsible for the
optical recombination of an exciton with momentum ${\bf p}$ accompanied by
simultaneous emission or absorption of a phonon.
The other designations are defined in the text after Hamiltonian
(\ref{X}) in Sec.\ref{IMP}.

Let us consider the two-photon emission accompanied by the coherent
two-exciton recombination and the emission of two phonons.\cite{2} These
phonons take away the excess momentum. Diagrams corresponding to such
two-photon emission are analogous to those shown in Fig.1. Only the  lines
corresponding to the interaction with impurities should be replaced by the
phonon lines responsible for the exciton-phonon interaction.

For the two-photon emission under consideration the energy conservation law
is

\begin{eqnarray}
\label{Eff}
2\Omega=\omega+\omega'+\omega^s_{-p-k'}+\omega^s_{p-k},
\end{eqnarray}
where $\omega^s_{p-k}$ and $\omega^s_{-p-k'}$ are the energies of the emitted
phonons. The total momentum of these phonons equals to the excess momentum
$\delta{\bf k}=-{\bf k}-{\bf k'}$.

In designations of Hamiltonian (\ref{HL}) the matrix element for the
two-photon emission under consideration is given by the formula analogous to
Eq.(\ref{Mij}):

\begin{eqnarray}
\label{Mff}
&&(\hat H_L)_{fi}=
\left[\frac{L^>_{-pk'}L^>_{pk}}{
\Omega-\omega^s_{p-k}-\epsilon_p-\omega+i\Gamma_p/2}+
\frac{L^>_{pk}L^>_{-pk'}}{
\Omega-\omega^s_{-p-k'}-\epsilon_p-\omega'+i\Gamma_p/2}\right]\times
\nonumber \\ && \times
\left(N_0\delta_p+(1-\delta_p)u_pv_p\right)+
\nonumber \\ && +
\left[\frac{L^>_{p-q,k'}L^>_{-p+q,k}}{
\Omega-\omega^s_{-p-k'}-\epsilon_{p-q}-\omega+i\Gamma_{p-q}/2}+
\frac{L^>_{-p+q,k}L^>_{p-q,k'}}{
\Omega-\omega^s_{p-k}-\epsilon_{p-q}-\omega'+i\Gamma_{p-q}/2}\right]\times
\nonumber \\ && \times
\left(N_0\delta_{p-q}+(1-\delta_{p-q})u_{p-q}v_{p-q}\right),
\end{eqnarray}
where $|i\rangle=|0\rangle_{exc}|0\rangle_{phot}|0\rangle_{phon}$;
$|f\rangle=
|0\rangle_{exc}|1_k,1_{k'}\rangle_{phot}|1_{p-k},1_{-p-k'}\rangle_{phon}$.

As in case of impurity-assisted two-photon emission,
using energy conservation law (\ref{Eff}) yields matrix element
(\ref{Mff}) expressed in terms of the anomalous Green functions of Bose
condensed excitons $\hat G_p(\omega)$ (see Eq.(\ref{G})) and the function
$N_0(\omega)$ (see Eq.(\ref{N0})):

\begin{eqnarray}
\label{ff}
&&(\hat H_L)_{fi}=-\left[2\pi iN_0(\Omega-\omega^s_{p-k}-\omega)\delta_p-
(1-\delta_p) \hat G_p(\Omega-\omega^s_{p-k}-\omega)\right]
L^>_{-pk'}L^>_{pk}-
\nonumber \\ && -
\left[2\pi iN_0(\Omega-\omega^s_{-p-k'}-\omega)\delta_{p-q}-
(1-\delta_{p-q})\hat G_{p-q}(\Omega-\omega^s_{-p-k'}-\omega)\right]
L^>_{p-q,k'}L^>_{-p+q,k}.
\nonumber \\
\end{eqnarray}

We shall limit ourselves to the consideration of the induced two-photon
emission accompanied by the coherent two-exciton recombination and
the emission of two {\it optical} phonons with negligibly small dispersion
($\omega^s_q=\omega^s_0$). The induced two-photon emission of this type
leads to the appearance of the spectral line on frequency
$2(\Omega-\omega^s_0)-\omega$, where $\omega$ is the frequency of the
incident light.\cite{n} This line is the phonon replica of spectral line
$2\Omega-\omega$ corresponding to the induced impurity-assisted
two-photon emission with the coherent two-exciton recombination.

For the differential cross-section of the induced two-photon emission
accompanied by the coherent two-exciton recombination and the emission
of two optical phonons one has

\begin{eqnarray}
\label{ds}
d\sigma^L=\frac{\omega(2\Omega_--\omega)^3}{2c^4}
\sum_{p}|(s_p)_{nm}e'^*_ne^*_m|^2do',
\end{eqnarray}
where

\begin{eqnarray}
\label{s}
&&(s_p)_{nm}=\left[2\pi iN_0(\Omega_--\omega)\delta_p-
(1-\delta_p)\hat G_p(\Omega_--\omega)\right](f^>_{-pk'})_n(f^>_{pk})_m+
\nonumber \\ && +
\left[2\pi iN_0(\Omega_--\omega)\delta_{p-q}-
(1-\delta_{p-q})\hat G_{p-q}(\Omega_--\omega)\right]
(f^>_{p-q,k'})_n(f^>_{-p+q,k})_m.
\end{eqnarray}
Here and further $\Omega_-=\Omega-\omega^s_0$. The summation over all
possible ${\bf p}$ takes into account the emission of two phonons
with momenta ${\bf p-k}$ and ${\bf -p-k'}$ twice:
$(s_p)_{nm}=(s_{-p+q})_{nm}$. That is why factor "2" appears in the
denominator of Eq.(\ref{ds}).

Further calculations are analogous to those for the induced two-photon
emission with the excess momentum transferred to two different
impurities. The overall cross-section of the induced two-photon emission
with the coherent two-exciton recombination assisted by two emitted optical
phonons is given by the formula

\begin{eqnarray}
\label{sL}
\sigma^L=\frac{\omega(2\Omega_--\omega)^3 2^{5/2}m^{3/2}\mu^{1/2}}{
9c^4\alpha_L\Gamma_L}
\frac{(\sqrt{\alpha_L^2+1}-1)^{1/2}}{\sqrt{\alpha_L^2+1}}
{\bf f}^2(\omega_L){\bf f}^2(\omega'_L),
\end{eqnarray}
where ${\bf f}(\omega_L)=
\int{\bf f}^>(p_L,k)do_{p_L}/(4\pi)$;
${\bf f}(\omega'_L)=\int{\bf f}^>(p_L,k')do_{p_L}/(4\pi)$; momentum
$p_L$ is defined by the condition $\epsilon(p_L)=|\Omega_--\omega|$.
Here $\alpha_L=|\Omega_--\omega|/\mu$, $\Gamma_L$ is the inverse lifetime of
the quasiparticle with energy $\epsilon(p_L)$ in the Bose-condensed exciton
system.

The induced two-photon emission accompanied by the coherent two-exciton
recombination and the transmission of the excess momentum to both a phonon
and an impurity can be considered analogously. The energy conservation
law for such two-photon emission is

\begin{eqnarray}
\label{2xj-e}
2\Omega=\omega+\omega'+\omega^s_{-p-k'},
\end{eqnarray}
where $\omega^s_{-p-k'}$ is the energy of the emitted phonon with momentum
${\bf -p-k'}$. The rest of the excess momentum ${\bf p-k}$ is transmitted to
an impurity.

The matrix element for the two-photon emission under consideration is

\begin{eqnarray}
\label{Mdf}
&&(\hat H_X+\hat H_L)_{fi}=
\left[\frac{L^>_{-pk'}X^i_{pk}}{\Omega-\epsilon_p-\omega+i\Gamma_p/2}+
\frac{X^i_{pk}L^>_{-pk'}}{
\Omega-\omega^s_{-p-k'}-\epsilon_p-\omega'+i\Gamma_p/2}\right]\times
\nonumber \\ && \times
\left(N_0\delta_p+(1-\delta_p)u_pv_p\right)+
\nonumber \\ && +
\left[\frac{X^i_{p-q,k'}L^>_{-p+q,k}}{
\Omega-\omega^s_{-p-k'}-\epsilon_{p-q}-\omega+i\Gamma_{p-q}/2}+
\frac{L^>_{-p+q,k}X^i_{p-q,k'}}{
\Omega-\epsilon_{p-q}-\omega'+i\Gamma_{p-q}/2}\right]\times
\nonumber \\ && \times
\left(N_0\delta_{p-q}+(1-\delta_{p-q})u_{p-q}v_{p-q}\right),
\end{eqnarray}
where $|i\rangle=|0\rangle_{exc}|0\rangle_{phot}|0\rangle_{phon}$;
$|f\rangle=|0\rangle_{exc}|1_k,1_{k'}\rangle_{phot}|1_{-p-k'}\rangle_{phon}$.

Using energy conservation law (\ref{2xj-e}), matrix element (\ref{Mdf}) can
be expressed in terms of the anomalous Green functions of Bose-condensed
excitons:

\begin{eqnarray}
\label{df}
&&(\hat H_X+\hat H_L)_{fi}=-\left[2\pi iN_0(\Omega-\omega)\delta_p+
(1-\delta_p)\hat G_p(\Omega-\omega)\right]L^>_{-pk'}X^i_{pk}-
\nonumber \\ &&
\left[2\pi iN_0(\Omega-\omega^s_{-p-k'}-\omega)\delta_{p-q}+
(1-\delta_{p-q})\hat G_{p-q}(\Omega-\omega^s_{-p-k'}-\omega)\right]
X^i_{p-q,k'}L^>_{-p+q,k}.
\nonumber \\
\end{eqnarray}

If the emitted phonon is optical ($\omega^s_{q}=\omega^s_0$),
the induced two-photon emission under consideration leads to the appearance
of the spectral line on frequency $2\Omega-\omega^s_0-\omega$ (see
energy conservation law (\ref{2xj-e})). Like the line on frequency
$2\Omega_--\omega$, this line is the phonon replica of the spectral line
$2\Omega-\omega$.

The differential cross-section of the induced two-photon emission with the
coherent two-exciton recombination assisted by both a phonon and an impurity
is

\begin{eqnarray}
\label{dsXL}
d\sigma^{XL}=\frac{\omega(2\Omega-\omega^s_0-\omega)^3}{c^4}
\sum_{i}^{}\sum_{p}|(s^i_p)_{nm}e'^*_ne^*_m|^2do',
\end{eqnarray}
where

\begin{eqnarray}
\label{si}
&&(s^i_p)_{nm}=\left[2\pi iN_0(\Omega-\omega)\delta_p+
(1-\delta_p)\hat G_p(\Omega-\omega)\right](f^>_{-pk'})_n(d^i_{pk})_m+
\nonumber \\ && +
\left[2\pi iN_0(\Omega_--\omega)\delta_{p-q}+
(1-\delta_{p-q})\hat G_{p-q}(\Omega_--\omega)\right]
(d^i_{p-q,k'})_n(f^>_{-p+q,k})_m.
\end{eqnarray}

The overall cross-section is

\begin{eqnarray}
\label{sXL}
\sigma^{XL}=
\frac{\omega(2\Omega-\omega^s_0-\omega)^32^{3/2}m^{3/2}\mu^{1/2}}{9c^4}
\left[
\frac{\left(\sqrt{\alpha^2_X+1}-1\right)^{1/2}}{\alpha_X\Gamma_X
\sqrt{\alpha^2+1}}
N{\bf d}^2(\omega_X){\bf f}^2(\omega'_X)+
\right. \nonumber \\ \left. +
\frac{(\sqrt{\alpha_L^2+1}-1)^{1/2}}{\alpha_L\Gamma_L\sqrt{\alpha_L^2+1}}
N{\bf d}^2(\omega'_L){\bf f}^2(\omega_L)\right],
\end{eqnarray}
where ${\bf f}(\omega'_X)=\int{\bf f}^>(p_X,k')do_{p_X}/(4\pi)$;
${\bf d}(\omega'_L)=\int{\bf d}(p_L,k')do_{p_L}/(4\pi)$.

\section{Raman scattering}
\label{RAMAN}
\ \ Not only two-photon emission but also Raman scattering
can be accompanied by the coherent two-exciton recombination. Initial and
final states of the exciton-photon system are
$|i\rangle=|0\rangle_{exc}|1_k\rangle_{phot}$;
$|f\rangle=|0\rangle_{exc}|1_{k'}\rangle_{phot}$ in this case. The energy
conservation law for the Raman scattering is

\begin{eqnarray}
\label{CRS}
\omega+(2\Omega-n\omega^s_0)=\omega',
\end{eqnarray}
where $n$ is an integer number. The excess momentum corresponding to the
scattering is taken by impurities ($n=0$), or by both the emitted optical
phonon and an impurity ($n=1$), or by two emitted optical phonons ($n=2$).
In general, an arbitrary number of photons $n$ can be emitted.
The Raman scattering with the coherent two-exciton recombination leads to the
appearance of anti-Stokes components on frequencies defined by Eq.(\ref{CRS}).

Raman scattering can be accompanied not only by the coherent
two-exciton recombination but also by the coherent two-exciton
{\it creation}. For the last type of Raman scattering the energy
conservation law is

\begin{eqnarray}
\omega-(2\Omega-n\omega^s_0)=\omega'.
\end{eqnarray}
It is easy to see that Raman scattering with the coherent two-exciton
creation and the transmission of the excess momentum to impurities
($n=0$) is possible only if $\omega>2\Omega$. Analogously, if the
excess momentum is taken by both the emitted optical phonon and an
impurity ($n=1$), or by two emitted optical phonons ($n=2$), Raman
light scattering  with coherent two-exciton creation takes place if
$\omega>2\Omega-\omega^s_0$ and $\omega>2\Omega_-$ correspondingly.

The consideration of Raman scattering is analogous to that
of the induced two-photon emission. The cross-sections of Raman
scattering accompanied by processes of the coherent two-exciton recombination
or creation can be obtained from the formulae of Sec.\ref{IMP},\ref{PHN}
by the appropriate substitutions. To shorten the text, we only point out
these substitutions below.

1) The cross-section of the impurity-assisted Raman scattering
accompanied by the
coherent two-exciton recombination can be obtained from Eq.(\ref{sX2}) (two
different impurities take the excess momentum) and Eq.(\ref{sX1}) (the
excess momentum is transferred to one of the impurities as a whole) by
replacing ${\bf d}(\omega_X)\rightarrow {\bf d}'(\widetilde{\omega}_X)$;
${\bf d}(\omega'_X)\rightarrow {\bf d}(\widetilde{\omega}'_X)$;
$\omega\rightarrow-\omega$, $\alpha_X\rightarrow\widetilde{\alpha}_X$,
$\Gamma_X\rightarrow\widetilde{\Gamma}_X$. Here
$\widetilde{\alpha}_X=(\Omega+\omega)/\mu$;
$\widetilde{\Gamma}_X$ is the inverse lifetime of the quasiparticle with
energy
$\epsilon(\widetilde{p}_X)=\Omega+\omega$ in the Bose-condensed exciton
system, ${\bf d}'(\widetilde{\omega}_X)=\int {\bf d}'(\widetilde{p}_X,k)
do_{\widetilde{p}_X}/(4\pi)$;
${\bf d}(\widetilde{\omega}'_X)=\int {\bf d}(\widetilde{p}_X,k')
do_{\widetilde{p}_X}/(4\pi)$.

2) The cross-section of Raman scattering with the coherent
two-exciton recombination assisted by two optical phonons can be obtained
from Eq. (\ref{sL}) by replacing
${\bf f}(\omega_L)\rightarrow {\bf f}'(\widetilde{\omega}_L)$;
${\bf f}(\omega'_L)\rightarrow {\bf f}(\widetilde{\omega}'_L)$;
$\omega\rightarrow -\omega$, $\alpha_L\rightarrow\widetilde{\alpha}_L$,
$\Gamma_L\rightarrow\widetilde{\Gamma}_L$. Here
$\widetilde{\alpha}_L=(\Omega_-+\omega)/\mu$;
$\widetilde{\Gamma}_L$ is the inverse lifetime of the quasiparticle with the
energy
$\epsilon(\widetilde{p}_L)=\Omega_-+\omega$ in Bose-condensed exciton system,
${\bf f}'(\widetilde{\omega}_L)=\int {\bf f}'^>(\widetilde{p}_L,k)
do_{\widetilde{p}_L}/(4\pi)$;
${\bf f}(\widetilde{\omega}'_L)=\int {\bf f}^>(\widetilde{p}_L,k')
do_{\widetilde{p}_L}/(4\pi)$.
The cross-section of Raman scattering with coherent two-exciton
recombination and the transmission of the excess momentum to the emitted
optical phonon and an impurity can be obtained from Eq.(\ref{sXL}) by
replacing
${\bf d}(\omega_X)\rightarrow {\bf d}'(\widetilde{\omega}_X)$;
${\bf f}(\omega'_X)\rightarrow {\bf f}(\widetilde{\omega}'_X)$;
${\bf f}(\omega_L)\rightarrow {\bf f}'(\widetilde{\omega}_L)$;
${\bf d}(\omega'_L)\rightarrow {\bf d}(\widetilde{\omega}'_L)$;
$\omega\rightarrow -\omega$; $\alpha_{X(L)}\rightarrow\tilde{\alpha}_{X(L)}$;
$\Gamma_{X(L)}\rightarrow\widetilde{\Gamma}_{X(L)}$. Here
${\bf f}(\widetilde{\omega}'_X)=\int {\bf f}^>(\widetilde{p}_X,k')
do_{\widetilde{p}_X}/(4\pi)$;
${\bf d}(\widetilde{\omega}'_L)=\int {\bf d}(\widetilde{p}_L,k')
do_{\widetilde{p}_L}/(4\pi)$.

3) The cross-section of the impurity-assisted Raman scattering with
the coherent two-exciton creation ($\omega>2\Omega$) can be obtained from
Eq.(\ref{sX2}) (two different impurities take the excess momentum)
and from Eq.(\ref{sX1}) (the excess momentum is transferred to one of the
impurities as a whole) by replacing
${\bf d}(\omega_X)\rightarrow {\bf d}'(\omega_X)$.

4) The cross-section of Raman scattering accompanied by the
coherent two-exciton creation and the emission of two optical phonons
$(\omega>2\Omega_-)$ can be obtained from Eq.(\ref{sL}) by replacing
${\bf f}(\omega_L)\rightarrow {\bf f}'(\omega_L)$.
The cross-section of Raman scattering with the coherent
two-exciton creation and the transmission of the excess momentum to the
emitted optical phonon and an impurity $(\omega>2\Omega-\omega^s_0)$
can be obtained from Eq.(\ref{sXL}) by replacing
${\bf d}(\omega_X)\rightarrow {\bf d}'(\omega_X)$;
${\bf f}(\omega_L)\rightarrow {\bf f}'(\omega_L)$.

\section{On possibility of experimental observation of
two-photon processes accompanied by coherent two-exciton recombination
(creation)}

\ \  In this section we shall analyze the possibility of the experimental
observation of the induced two-photon emission and Raman scattering
accompanied by the coherent two-exciton recombination or creation. As an
example, the induced two-photon emission assisted by optical phonons
will be considered.

The intensity of the light $I_L(2\Omega_--\omega)$ on frequency
$2\Omega_--\omega$ corresponding to the induced two-photon emission
under consideration  is given by the expression

\begin{eqnarray}
\label{IL}
I_L(2\Omega_--\omega)=\frac{2\Omega_--\omega}{\omega}
\sigma^L(\omega)I(\omega),
\end{eqnarray}
where $I(\omega)(W/cm^2)$ is the intensity of the incident light on frequency
$\omega$, $\sigma^L(\omega)$ is the cross-section of the induced
two-photon emission assisted by optical phonons (see Eq.(\ref{sL})).

If the frequency of the incident light $\omega<\Omega_-$, the spectral line
corresponding to the induced two-photon emission under consideration
will be observed on frequency $2\Omega_--\omega>\Omega_-$. At $T=0$ this line
is out of the spectral interval of the luminescence of the Bose-condensed
excitons assisted by optical phonons which takes place on frequencies
$\omega'<\Omega_-$ at zero temperature (see Appendix \ref{B}). If
$\omega>\Omega_-$, more detailed analyses is needed, because the light
corresponding to the induced two-photon emission under consideration will
be observed against the background of the luminescence of the Bose-condensed
excitons assisted by optical phonons. Intensity (\ref{IL}) can be
represented as a sum of two terms

\begin{eqnarray}
I^L(2\Omega_--\omega)=\Delta I^L(2\Omega_--\omega)+I^L_r(2\Omega_--\omega),
\end{eqnarray}
where $I^L_r(2\Omega_--\omega)$ is the intensity of the induced
{\it two-stage} two-photon emission consisting of two successive
processes {\it each} of which satisfies the energy conservation law. If the
frequency of the incident light $\omega>\Omega_-$, the induced two-stage
two-photon emission is the result of

{\bf I.}  The {\it spontaneous} recombination of an exciton with momentum
${\bf p}_L$ accompanied by appearance of the quasiparticle with energy
$\epsilon(p_L)=\omega-\Omega_-$ in the exciton system and spontaneous
emission of a photon on frequency $2\Omega_--\omega$.

{\bf II.} The {\it induced} recombination of an exciton with momentum
${\bf -p}_L$ accompanied by disappearance of a quasiparticle with energy
$\epsilon(p_L)$ and induced emission of photon $\omega$.

The number of quasiparticles with energy $\epsilon(p_L)=\omega-\Omega_-$
appearing in the exciton system per second due to the
spontaneous exciton
recombination is $I^L_s(2\Omega_--\omega)/(2\Omega_--\omega)$ (here
$I^L_s(\omega)$ is the intensity of the luminescence, see Appendix \ref{B}
and also Ref.\cite{GRIFFIN94}). These quasiparticles disappear during
their effective lifetime $\tau^L$. The disappearance of some part of the
quasiparticles is accompanied by the induced emission on frequency
$\omega$. Therefore, if $\omega>\Omega_-$, the intensity
$I^L_r(2\Omega_--\omega)$ can be found from the equation

\begin{eqnarray}
\label{ILR}
I^L_r(2\Omega_--\omega)=\frac{\tau^L}{\tau^L_r}I^L_s(2\Omega_--\omega),
\end{eqnarray}
where $\tau^L_r$ is the lifetime of the quasiparticle $\epsilon(p_L)$ with
respect to its disappearance accompanied by the induced emission on frequency
$\omega$.

The ratio $\tau^L/\tau^L_r$ defines the portion of
quasiparticles with energies $\epsilon(p_L)$, whose disappearance is
accompanied by the induced emission on frequency $\omega$.
Eq.(\ref{ILR}) gives the {\it part} of the {\it total} intensity of the
luminescence on frequency $2\Omega_--\omega$ corresponding to the induced
two-stage two-photon emission. Thus, if the frequency of the incident light
$\omega>\Omega_-$, the intensity of the spectral line on frequency
$2\Omega_--\omega$ corresponding to the induced two-photon emission
exceeds the intensity of the luminescence on this frequency by
$\Delta I^L(2\Omega_--\omega)$.

Using the Fermi's golden rule yields

\begin{eqnarray}
\label{tr}
\frac{1}{\tau^L_r}=\frac{(2\pi)^2}{3c}{\bf f}^2(\omega_L)u^2_{p_L}
I(\omega),
\nonumber \\
u^2_{p_L}=\frac{1}{2}(1+\sqrt{\alpha^2_L+1}/\alpha_L),
\end{eqnarray}
where $u_{p_L}$ is the coefficient of the Bogoliubov transformation
corresponding to momentum $p_L$.

Inserting Eqs.(\ref{tr}) and (\ref{Isi}) (see Appendix \ref{B}) into
Eq.(\ref{ILR}) one has $I^L_r(2\Omega_--\omega)=I^L(2\Omega_--\omega)/2$.
Therefore, the intensity of the spectral line on frequency
$2\Omega_--\omega$ corresponding to the induced two-photon emission exceeds
the luminescence intensity on this frequency by the quantity

\begin{eqnarray}
\label{DIL}
\Delta I^L(2\Omega_--\omega)=
\frac{(2\Omega_--\omega)^4 2^{3/2}m^{3/2}\mu^{1/2}}{9c^4\alpha_L\Gamma_L}
\frac{(\sqrt{\alpha_L^2+1}-1)^{1/2}}{\sqrt{\alpha_L^2+1}}
{\bf f}^2(\omega_L){\bf f}^2(\omega'_L).
\end{eqnarray}

Using Eq.(\ref{sL}) we shall estimate cross-section $\sigma^L$ of the
induced two-photon emission with coherent two-exciton recombination assisted
by optical phonons. The cross-section expressed in ordinary units is

\begin{eqnarray}
\label{CGS}
\sigma^L=\tau^LV
\frac{\omega(2\Omega_--\omega)^32^{5/2}m^{3/2}\mu^{1/2}}{9c^4\hbar^4}
\frac{(\sqrt{\alpha_L^2+1}-1)^{1/2}}{\alpha_L\sqrt{\alpha^2_L+1}}
{\bf f}^2(\omega_L){\bf f}^2(\omega'_L),
\end{eqnarray}
where $V$ is the volume of excitons interacting with the incident light,
$\tau^L$ is the lifetime of a quasiparticle with energy
$\hbar(\Omega_--\omega)$ in the Bose-condensed exciton system,
$\alpha_L=\hbar|\Omega_--\omega|/\mu$.

There are the reports about the experimental observation of Bose condensation
of excitons in $Cu_2O$.\cite{PARA} To estimate the cross-section
we consider the Bose-condensed excitons of density $n=10^{19} cm^{-3}$ in
the $Cu_2O$ at $T=0$. The exciton mass is equal to $m=2.7m_e$ in this
crystal, the exciton radius is $a=7\AA$, the energy corresponding to the
recombination of an exciton with $p=0$ is $\hbar\Omega\simeq 2eV$. The optical
recombination accompanied by the emission of an optical phonon with energy
$\hbar\omega^s_0\simeq 10meV$ is typical for excitons in $Cu_2O$.

Chemical potential $\mu$ of excitons can be roughly estimated from the
formula for the chemical potential of weakly interacting Bose gas:

\begin{eqnarray*}
\mu=\frac{4\pi\hbar^2}{m}na\simeq 2(meV)
\end{eqnarray*}
at exciton density $n=10^{19}cm^{-3}$.

In experiment \cite{PARA} the excitons were pumped by powerful nanosecond
laser pulses (the wavelength $\lambda\simeq 500 nm$) focused in the spot of
diameter $d\simeq 30 \mu m$ on the crystal surface. Thus, the volume of
Bose-condensed excitons can be estimated by the formula $V=d^2l$, where
$l\simeq 1\mu m$ is the penetration depth of the radiation with the
wavelength $500nm$.

If the frequency of the incident light $\omega\rightarrow\Omega_-$
($\alpha_L\rightarrow 0$), cross-section $\sigma^L$ increases.
We shall suppose that $\hbar(\Omega_--\omega)=\mu$. In this case
${\bf f}(\omega_L)\simeq {\bf f}(\omega'_L)\simeq {\bf F}(\omega_L)$, where
${\bf F}(\omega_L)=\int {\bf F}^>(p_L,k)do_{p_L}/(4\pi)$ is the matrix
element of the optical recombination of an isolated exciton assisted by an
optical phonon (see Appendix \ref{D}). ${\bf F}(\omega_L)$ can be estimated
from the expression

\begin{eqnarray}
\frac{1}{\tau_{exc}}=\frac{4\Omega^3_-}{3c^3\hbar}{\bf F}^2(\omega_L),
\end{eqnarray}
where $\tau_{exc}$ is the lifetime of an exciton with respect to
the spontaneous recombination accompanied by the emission of a photon with
energy $\hbar\Omega_-$ and the optical phonon with energy $\hbar\omega^s_0$.
For paraexcitons in $Cu_2O$ the lifetime is $\tau_{exc}\sim100 \mu s$
(see Chap.13 in Ref.\cite{BOSE}).

The lifetime of the quasiparticle in the Bose-condensed exciton system
$\tau^L$ is the subject of further investigation. It can be sufficiently
smaller than the radiative lifetime of the excitons $\tau_{exc}$ even at
$T=0$ due to the possibility of the quasiparticle scattering with the
emission of a phonon. Supposing $\tau^L$ is in the range
$10^{-11}$--$10^{-5}sec$ (the lower boundary corresponds to the condition
$\Gamma_L=10^{-1}\epsilon(p_L)$, the upper one is $10^{-1}\tau_{exc}$)
yields the estimation $\sigma^L=10^{-16}$--$10^{-10}cm^2$.

The radiative lifetime of orthoexcitons in $Cu_2O$ is about $300ns$.
Supposing $\tau^L$ is in the range $10^{-11}$--$10^{-9}sec$ (the upper
boundary is the lifetime of an orthoexciton with respect to ortho-to-para
phonon-assisted conversion in this case) yields
$\sigma^L=10^{-11}$--$10^{-9}cm^2$. So the induced two-photon emission
accompanied by the coherent two-exciton recombination possibly can be
observed in $Cu_2O$.

The cross-section of Raman scattering assisted by two emitted optical
phonons is quadratic in the product of matrix elements
${\bf f}'(\widetilde{\omega}_L)=\int {\bf f}'^>(\widetilde{p}_L,k)
do_{\widetilde{p}_L}/(4\pi)$ and
${\bf f}(\widetilde{\omega}'_L)=\int {\bf f}^>(\widetilde{p}_L,k')
do_{\widetilde{p}_L}/(4\pi)$, where $\widetilde{p}_L$ is defined by the
condition $\epsilon(\widetilde{p}_L)=|\omega+\Omega_-|$
(see Sec.\ref{RAMAN}). Since the band gap in $Cu_2O$ is wide, the relation
$\epsilon(\widetilde{p}_L)\gg\omega^s_0$ is valid
($\Omega_-\sim10^2\omega^s_0$ in $Cu_2O$). In this case
${\bf f}(\widetilde{\omega}'_L)$ and ${\bf f}'(\widetilde{\omega}_L)$
are negligibly smaller than ${\bf f}(\omega_L)$ involved in the cross-section
of the induced two-phonon emission at $|\Omega_--\omega|\sim\mu$ (see
Eqs. (\ref{L}),(\ref{L'}) in Appendix \ref{D}). Besides, the cross-section
of Raman scattering under consideration is proportional to the lifetime
of a quasiparticle with energy $\epsilon(\widetilde{p}_L)=|\omega+\Omega_-|$
which is supposed to be essentially smaller than $\tau^L$ of a quasiparticle
with energy $\epsilon(p_L)=\mu$ involved in the cross-section (\ref{CGS})
at $|\Omega_--\omega|=\mu$. Therefore, the experimental observation
of Raman scattering accompanied by the coherent  two-exciton recombination
is hardly possible in $Cu_2O$, contrary to the induced two-photon emission
(see above). As for Raman scattering accompanied by the coherent two-exciton
creation the situation is analogous.

\section{Conclusion}

We have shown that the induced two-photon emission and Raman scattering
accompanied by the recombination (or creation) of two excitons with
opposite momenta leaving the exciton occupation numbers with ${\bf p}\neq 0$
unchanged takes place
in the interacting Bose-condensed exciton system. The excess momentum can be
transferred to impurities or phonons involved in these processes. Both the
induced two-photon emission and Raman scattering under consideration
can be used to probe exciton Bose condensation, because they are absent if
excitons are in normal state. The recombination (creation) of two excitons
with opposite momenta leaving the exciton occupation numbers unchanged is
called coherent two-exciton recombination (creation) in the paper.

If the frequency of the incident light $\omega<2\Omega$ ($\Omega$ is the
frequency corresponding to the recombination of an exciton with ${\bf p}=0$),
there is the spectral line on frequency $2\Omega-\omega$ corresponding to
the induced impurity-assisted two-photon emission accompanied by the coherent
two-exciton recombination. The anti-Stokes line on frequency $\omega+2\Omega$
corresponding to the impurity-assisted Raman scattering accompanied by
the coherent two-exciton recombination also appears. If $\omega>2\Omega$,
there are both Stokes and anti-Stokes lines on frequencies $\omega\pm2\Omega$
appear. The Stokes line corresponds to the coherent two-exciton creation.
The induced two-photon emission is impossible in this case.

Spectral lines $|\omega\pm2\Omega|$ have phonon replicas on frequencies
$|\omega\pm(2\Omega-n\omega^s_0|$ corresponding to the transmission of the
excess momentum (partially or as a whole) to optical phonons of frequency
$\omega^s_0$ ($n$ is an integer number). The quantitative estimation shows
that the spectral line on frequency $2(\Omega-\omega^s_0)-\omega$
corresponding to the induced phonon-assisted two-photon emission can be
experimentally observed in $Cu_2O$.

\section*{Acknowledgment}

The work is supported by Russian Foundation of Basic Research, INTAS and
by Program "Fundamental Spectroscopy".

\appendix

\section{Effective matrix elements for the optical exciton recombination}
\label{D}

\ \ \ Hamiltonian responsible for the interaction of excitons with
impurities, phonons and electromagnetic field is

\begin{eqnarray}
&&\hat V=\hat U+\hat W+\hat D, \nonumber \\ &&
\hat U=\sum_{jpq}^{}U^j_{qp}Q^+_qQ_p, \nonumber \\ &&
\hat W=\sum_{pq}^{}\left(W_{qp}Q^+_qQ_pb_{q-p}+W^*_{pq}Q^+_qQ_pb^+_{p-q}
\right), \nonumber \\ &&
\hat D=\sum_{q}^{}\left(D_qQ_qc^+_q+D'_qQ_{-q}c_q+h.c.\right),
\end{eqnarray}
where Hamiltonian $\hat U$ corresponds to scattering of excitons on
impurities, $\hat W$ and $\hat D$ are Hamiltonians responsible for
exciton-phonon and exciton-photon interactions correspondingly;
$D_q=i\sqrt{2\pi\omega_q}({\bf e^*d}_q)$,
$D'_q=-i\sqrt{2\pi\omega_q}({\bf ed}_q)$.

In the first order of perturbation series on $\hat V$ only the direct
optical recombination (creation) of an exciton is possible. In the second
order of perturbation series on $\hat V$ the indirect optical recombination
(creation) of an exciton is also possible. For example, matrix element
$X^j_{pq}$ responsible for the optical recombination of an exciton with the
transmission of the excess momentum to impurity $j$ (see
Hamiltonian (\ref{X})) can be obtained from the expression

\begin{eqnarray}
&&X^j_{pq}\langle f|Q_pc^+_q|i\rangle=
\nonumber \\ && =
\frac{\langle f|D_qQ_qc^+_q|\nu_1\rangle\langle\nu_1|U^j_{qp}Q^+_qQ_p
|i\rangle}{E_i-E_{\nu_1}}+
\frac{\langle f|U^j_{qp}Q^+_qQ_p|\nu_2\rangle\langle\nu_2|D_qQ_qc^+_q
|i\rangle}{E_i-E_{\nu_2}},
\nonumber \\
\end{eqnarray}
where $|i\rangle=|..m_p..\rangle_{exc}|..n_q..\rangle_{phot}$
and $|f\rangle=|..m_p-1..\rangle_{exc}|..n_q+1..\rangle_{phot}$ are
initial and final states of the exciton-photon system,
$|\nu_1\rangle=|..m_q+1;m_p-1..\rangle_{exc}|..n_q..\rangle_{phot}$ and
$|\nu_2\rangle=|..m_q-1;m_p..\rangle_{exc}|..n_q+1..\rangle_{phot}$ are
its intermediate states, $E_{\nu_1}$ and $E_{\nu_2}$ are the energies
of the intermediate states. Here $m_p$ is the occupation number of
excitons with momentum ${\bf p}$, $n_q$ is the photon occupation
number.

In case of the dilute Bose-condensed exciton gas at $T=0$, we have
$m_p=v^2_p$; $E_i-E_{\nu_1}=-\epsilon_p-\epsilon_q$;
$E_i-E_{\nu_2}=\Omega-\omega_q-\epsilon_q$, where $v_p$ is the coefficient
of Bogoliubov transformation (\ref{uv}), $\epsilon_p$ is
the energy of the Bogoliubov quasiparticle in the exciton system. Therefore,

\begin{eqnarray}
X^j_{pq}=i\sqrt{2\pi\omega_q}({\bf e^*d}^j_{pq}); \
{\bf d}^j_{pq}=-{\bf d}_qU^j_{qp}\left[\frac{u^2_q}{\epsilon_p+\epsilon_q}-
\frac{v^2_q}{\Omega-\omega_q-\epsilon_q}\right].
\end{eqnarray}

Matrix element $X'^j_{pq}$ (see Hamiltonian (\ref{X})) is given by the
analogous expression:

\begin{eqnarray}
X'^j_{pq}=-i\sqrt{2\pi\omega_q}({\bf ed'}^j_{pq}); \
{\bf d'}^j_{pq}=-{\bf d}_qU^j_{-qp}\left[\frac{u_q^2}{\epsilon_p+\epsilon_q}-
\frac{v^2_q}{\Omega+\omega_q-\epsilon_q}\right].
\end{eqnarray}

Matrix element $L^>_{pq}$ corresponding to the optical phonon-assisted
exciton \\
recombination (see Hamiltonian (\ref{HL})) can be obtained from
the expression

\begin{eqnarray}
\label{L}
&&L^>_{pq}\langle f|Q_pc^+_qb^+_{p-q}|i\rangle=
\frac{\langle f|D_qQ_qc^+_q|\nu_1\rangle
\langle\nu_1|W^*_{pq}Q^+_qQ_pb^+_{p-q}|i\rangle}{E_i-E_{\nu_1}}+
\nonumber \\ && +
\frac{\langle f|W^*_{pq}Q^+_qQ_pb^+_{p-q}|\nu_2\rangle
\langle\nu_2|D_qQ_qc^+_q|i\rangle}{E_i-E_{\nu_2}},
\end{eqnarray}
where \\
$|i\rangle=|..m_p..\rangle_{exc}|..s_{p-q}..\rangle_{phon}
|..n_q..\rangle_{phot}$ and
$|f\rangle=|..m_p-1..\rangle_{exc}|..s_{p-q}+1..\rangle_{phon}
|..n_q+1..\rangle_{phot}$ are initial and final states of the system
"excitons and phonons + electromagnetic field",
$|\nu_1\rangle=|..m_q+1;m_p-1..\rangle_{exc}
|..s_{p-q}+1..\rangle_{phon}|..n_q..\rangle_{phot}$ and
$|\nu_2\rangle=|..m_q-1;m_p..\rangle_{exc}
|..s_{p-q}..\rangle_{phon}|..n_q+1..\rangle_{phot}$ are the intermediate
states of this system.

Supposing the phonons to be optical and neglecting their dispersion, for the
dilute Bose-condensed excitons at $T=0$ one has

\begin{eqnarray}
\label{L'}
L^>_{pq}=i\sqrt{2\pi\omega_q}({\bf e^*f}^>_{pq}); \
{\bf f}^>_{pq}=-{\bf d}_qW^*_{pq}
\left[\frac{u^2_q}{\epsilon_p+\epsilon_q+\omega^s_0}-
\frac{v^2_q}{\Omega-\omega_q-\epsilon_q}\right].
\end{eqnarray}
The analogous expressions for the other matrix elements in Hamiltonian
(\ref{HL}) are

\begin{eqnarray}
&&L^<_{pq}=i\sqrt{2\pi\omega_q}({\bf e^*f}^<_{pq}); \
{\bf f}^<_{pq}=
-{\bf d}_qW_{qp}\left[\frac{u^2_q}{\epsilon_p+\epsilon_q-\omega^s_0}-
\frac{v^2_q}{\Omega-\omega_q-\epsilon_q}\right];
\nonumber \\ &&
L'^>_{pq}=-i\sqrt{2\pi\omega_q}({\bf ef'}^>_{pq}); \
{\bf f'}^>_{pq}=
-{\bf d'}_qW^*_{p,-q}\left[\frac{u^2_q}{\epsilon_p+\epsilon_q+\omega^s_0}-
\frac{v^2_q}{\Omega+\omega_q-\epsilon_q}\right];
\nonumber \\ &&
L'^<_{pq}=-i\sqrt{2\pi\omega_q}({\bf ef'}^<_{pq}); \
{\bf f'}^<_{pq}=
-{\bf d}_qW^*_{-qp}\left[\frac{u^2_q}{\epsilon_p+\epsilon_q-\omega^s_0}-
\frac{v^2_q}{\Omega+\omega_q-\epsilon_q}\right].
\end{eqnarray}

If $\epsilon_p+\epsilon_q\ll\omega^s_0$ and
$\Omega-\omega^s_0-\omega_q\ll\omega^s_0$, matrix element
${\bf f}^>_{pq}$ is

\begin{eqnarray}
\label{fF}
{\bf f}^>_{pq}=-\frac{{\bf d}_qW^*_{pq}}{\omega^s_0}.
\end{eqnarray}
In this case ${\bf f}^>_{pq}$ coincides with the matrix element for the
optical recombination of an isolated exciton assisted by an optical phonon.
In fact, using (\ref{L}) yields

\begin{eqnarray}
L^>_{pq}=i\sqrt{2\pi\omega_q}({\bf e^*F}^>_{pq}); \
{\bf F}^>_{pq}=\frac{{\bf d}_qW^*_{pq}}{E_p-E_q-\omega^s_0},
\end{eqnarray}
where $E_p=p^2/(2m)$. It is easy to see that ${\bf f}^>_{pq}={\bf F}^>_{pq}$
if $E_p-E_q\ll\omega^s_0$.

\section{Diagrams of two-photon processes accompanied by coherent
two-exciton recombination}
\label{A}

\ \ The induced two-photon emission and Raman scattering
accompanied by
the coherent two-exciton recombination (creation) can be considered in the
formalism of Green functions of Bose-condensed excitons. Diagrams shown
in Figs.1-2 are responsible for the corresponding elements of
$S$-matrix expressed in terms of  anomalous Green functions of
Bose-condensed excitons. Below it will be illustrated on the example of the
impurity-assisted two-photon emission accompanied by the coherent
two-exciton recombination.

In the Heisenberg representation Hamiltonian (\ref{X}) is

\begin{eqnarray}
\label{XINT}
\hat H_X(t)=\sum_{jpq}^{}\left(X^j_{pq}e^{-i\Omega t}
Q_p(t)c^+_{q}(t)+X'^j_{pq}e^{-i\Omega t}
Q_p(t)c_{q}(t)+h.c.\right),
\end{eqnarray}
where $c_q(t)=c_qe^{-i\omega_qt}$; $Q_p(t)=Q_pe^{-i\epsilon(p)t}$. The
exciton energy is measured from the bottom of the exciton band:
$\epsilon(0)=0$.

We shall expand the evolution operator
$\hat S(t)=T{\rm exp}(-i\int_{-\infty}^{t}\hat H_X(t')dt')$ as a power
series in $\hat H_X$ up to the second order, inclusive. The element of
$S$-matrix corresponding to the two-photon emission with the coherent
two-exciton recombination and the transmission of momentum ${\bf p-k}$
to impurity $i$ and momentum ${\bf -p-k'}$ to impurity $j$ is

\begin{eqnarray}
\label{Sij}
&&(S^{ij}_p)_{fi}=\frac{(-i)^2}{2!}\int\int_{-\infty}^{\infty}dt'dt''
e^{-i\Omega(t'+t'')}\times
\nonumber \\ && \times \left\{\left[
X^i_{pk}X^j_{-pk'}\left(N_0\delta_p+i(1-\delta_p)\hat G_{-p}(t'-t'')\right)+
\right. \right. \nonumber \\ && \left. \left. +
X^j_{-p+q,k}X^i_{p-q,k'}
\left(N_0\delta_{p-q}+i(1-\delta_{p-q})\hat G_{p-q}(t'-t'')\right)\right]
\langle f|c^+_{k}(t')c^+_{k'}(t'')|i\rangle_{phot}+
\right. \nonumber \\ && \left. + \left[
X^j_{-pk'}X^i_{pk}\left(N_0\delta_p+i(1-\delta_p)\hat G_p(t'-t'')\right)+
\right. \right. \nonumber \\ && \left. \left. + X^i_{p-q,k'}X^j_{-p+q,k}
\left(N_0\delta_{p-q}+i(1-\delta_{p-q})\hat G_{-p+q}(t'-t'')\right)\right]
\langle f|c^+_{k'}(t')c^+_{k}(t'')|i\rangle_{phot}\right\},
\nonumber \\
\end{eqnarray}
where $\hat G_p(t'-t'')=-i\langle TQ_{-p}(t')Q_p(t'')\rangle$ is the
anomalous Green function of the Bose-condensed excitons at $T=0$
(see, for example, Ref.\cite{AGD}).

The sum of diagrams shown in Fig.1 corresponds to the obtained element
of $S$-matrix. The anomalous Green function of the Bose-condensed excitons
is denoted by the thick line with oncoming arrows in this figure. If the
momenta of this line are zero, this line denotes factor $N_0$ of
macroscopic occupation of the exciton state with ${\bf p}=0$. The wavy
lines are responsible for the photon creation operators. The vertices on
these diagrams correspond to the matrix elements $X^i_{pk}$; $X^j_{pk}$,
where $p$ and $k$ are the momenta of the exciton and the photon lines
coming out of the vertex.

The element of $S$-matrix corresponding to the two-photon emission
accompanied by the coherent two-exciton recombination and the transmission
of the excess momentum to one of the impurities as a whole can be obtained
analogously:

\begin{eqnarray}
\label{Sjj}
&&(S^{jj})_{fi}=\frac{(-i)^2}{2}\int\int_{-\infty}^{\infty}dt'dt''
e^{-i\Omega(t'+t'')}\times
\nonumber \\ && \times
\sum_{p}^{}\left\{X^j_{-pk}X^j_{pk'}
\left[N_0\delta_p+i(1-\delta_p)\hat G_p(t'-t'')\right]
\langle f|c^+_{k}(t')c^+_{k'}(t'')|i\rangle_{phot}+
\right. \nonumber \\ && \left. +
X^j_{-pk'}X^j_{pk}
\left[N_0\delta_p+i(1-\delta_p)\hat G_p(t'-t'')\right]
\langle f|c^+_{k'}(t')c^+_{k}(t'')|i\rangle_{phot}\right\}.
\end{eqnarray}
The sum of diagrams with all possible momenta ${\bf p}$ shown in Fig.2
corresponds to this element of $S$-matrix.

Integrating on $t'-t''$ and $t''$ in Eqs.(\ref{Sij}), (\ref{Sjj}) yields
the formulae (\ref{Mij}), (\ref{1}) for the matrix elements
$(\hat H^{ij}_X)_{fi}$ and $(\hat H^{jj}_X)_{fi}$.

\section{One-photon emission by Bose-condensed excitons at $T=0$}

\label{B}

\ \ There are no quasiparticles in the Bose-condensed exciton system at
zero temperature. Therefore, the recombination of an exciton with momentum
${\bf p}\neq 0$ is inevitably accompanied by the creation of a quasiparticle
with momentum ${\bf -p}$ in the exciton system at $T=0$. The difference
between the energies of the exciton system before and after the recombination
is $\Omega-\epsilon_p$ in this case, where $\epsilon_p$ is the energy
of the created quasiparticle. Hence, the one-photon emission by the
Bose-condensed excitons with the excess momentum elastically transferred
to the impurities takes place on frequencies $\omega<\Omega$ at $T=0$.
Analogously, the frequency of the one-photon emission with the excess
momentum transferred to an optical phonon takes place at $\omega<\Omega_-$.

At $T=0$ the matrix element for the phonon-assisted exciton
recombination with the emission of photon $\omega$ is

\begin{eqnarray}
\label{M1}
&&(\hat L)_{fi}=i\sqrt{2\pi\omega_k} ({\bf e^*f}_{p_Lk})v_{p_L}\sqrt{n_k+1};
\nonumber \\ &&
v^2_{p_L}=\frac{\sqrt{\alpha_L^2+1}-\alpha_L}{2\alpha_L},
\end{eqnarray}
where $v_{p_L}$ is the coefficient of the Bogoliubov transformation
corresponding to the energy $\epsilon(p_L)=\Omega_--\omega$, $n_k$ is the
photon occupation number in the initial state,
$\alpha_L=(\Omega_--\omega)/\mu$. Here
$|i\rangle=|0\rangle_{exc}|0\rangle_{phon}|n_k\rangle_{phot}$  and
$|f\rangle=
|1_{-p_L}\rangle_{exc}|1_{p_L-k}\rangle_{phon}|n_k+1\rangle_{phot}$,
where $|1_{-p_L}\rangle_{exc}$ is the state of the exciton system containing
the quasiparticle with momentum ${\bf -p}_L$, $|0\rangle_{exc}$ is the
vacuum state with respect to quasiparticles in the exciton system.

Using the Fermi's golden rule yields the intensity of the one-photon
emission by the Bose-condensed excitons with the
excess momentum transferred to an optical phonon:

\begin{eqnarray}
\label{Isi}
&&I^L_{si}(\omega)=I^L_s(\omega)+I^L_i(\omega),
\nonumber \\ &&
I^L_s(\omega)=\frac{\omega^42^{1/2}m^{3/2}\mu^{1/2}}{3\pi^2c^3}
\frac{(\sqrt{\alpha_L^2+1}-1)^{1/2}{\bf f}^{2}(\omega_L)}{\sqrt{\alpha_L^2+1}
(\sqrt{\alpha^2_L+1}+\alpha_L)}\theta(\Omega_--\omega),
\nonumber \\ &&
I^L_i(\omega)=\frac{\omega}{c}
\frac{2^{1/2}m^{3/2}\mu^{1/2}(\sqrt{\alpha_L^2+1}-1)^{1/2}
{\bf f}^{2}(\omega_L)}{3\sqrt{\alpha_L^2+1}(\sqrt{\alpha_L^2+1}+\alpha_L)}
\theta(\Omega_--\omega)I(\omega),
\end{eqnarray}
where $\alpha_L=(\Omega_--\omega)/\mu$; $\theta(x)$ is the Heaviside
unit-step function. Here $I^L_s$, $I^L_i$ are the intensities of the
spontaneous and the induced emission, $I$ is the intensity of the incident
light.

Analogous formulae for the intensity of the one-photon emission with the
excess momentum elastically transferred to the impurity can be obtained from
Eqs.(\ref{Isi}) by replacing
${\bf f}(\omega_L)\rightarrow N{\bf d}(\omega_X)$;
$\Omega_-\rightarrow\Omega$; $\alpha_L\rightarrow\alpha_X$, where
$\alpha_X=(\Omega-\omega)/\mu$.

\newpage

\begin{center}
{\bf {\Large Caption for Figures }}
\end{center}

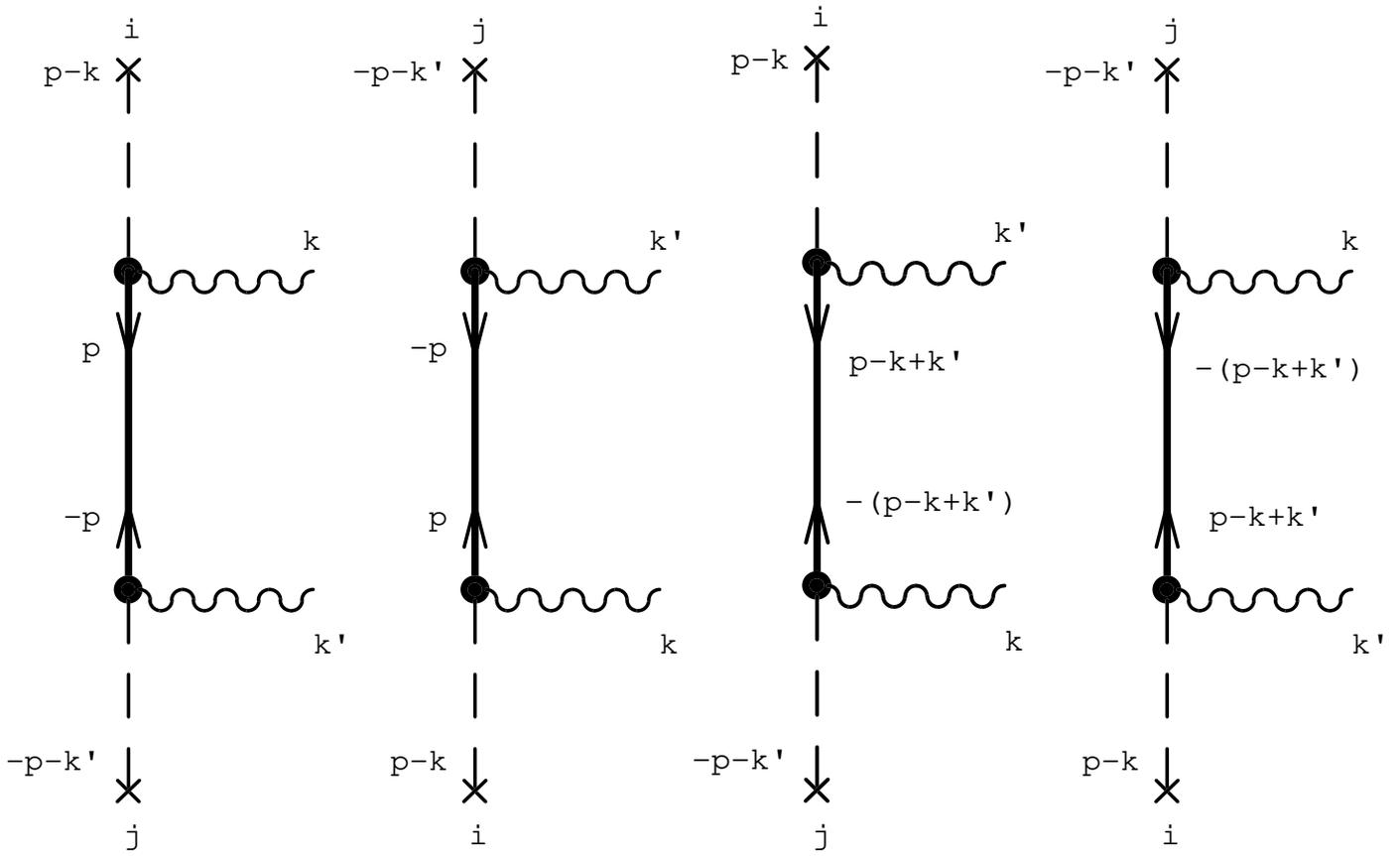
{\bf Fig.1.} Diagrams of the two-photon emission accompanied by the coherent
two-exciton recombination with the excess momentum transferred to {\it two
different} impurities.

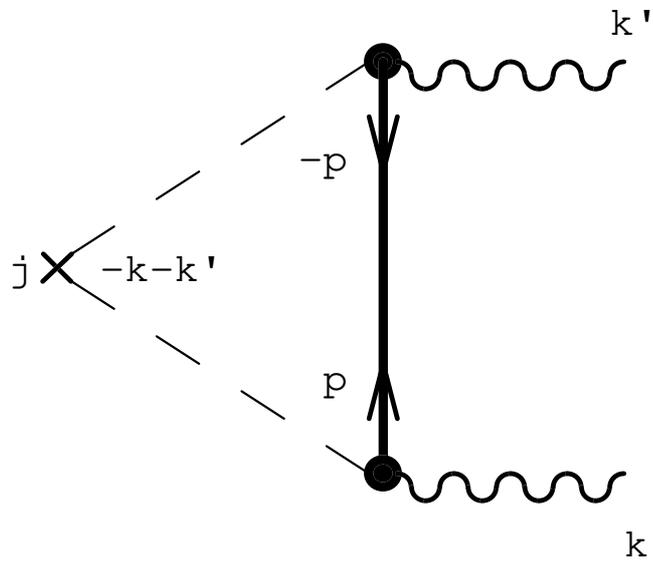
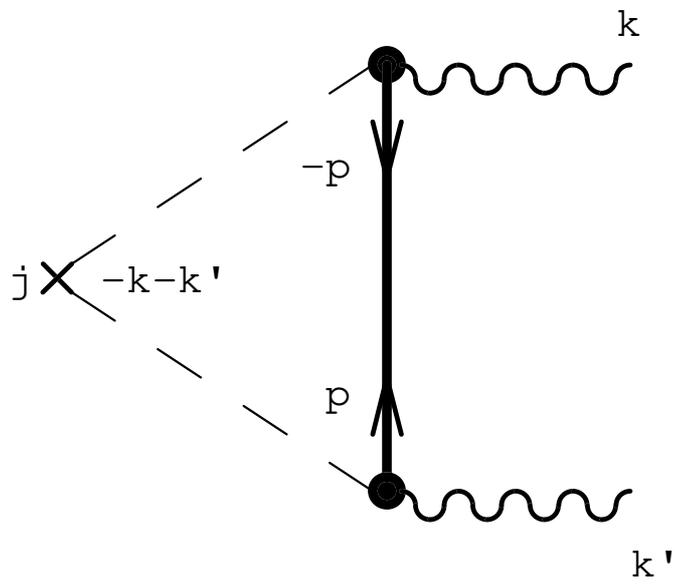
{\bf Fig.2.} Diagrams of the induced two-photon emission accompanied by
the coherent two-exciton recombination with the excess momentum transferred
to {\it one} of the impurities {\it as a whole}.


\begin{references}
\bibitem{GIMP} We suppose that the exciton scattering on impurities does not
essentially change the Green functions of excitons. It is possible if the
concentration of impurities is sufficiently low.
\bibitem{GP} $\Gamma_p$ is inversely proportional to the relaxation time of
a quasiparticle with momentum ${\bf p}$ in the exciton system. $\Gamma_p$
can be essentially greater than $\Gamma_0$ for the condensate exciton.
\bibitem{2} The analogous process with the absorption of phonons is
impossible at zero temperature, because phonons are absent at $T=0$.
\bibitem{n} In general, an arbitrary number of phonons can be involved in the
induced two-photon emission accompanied by the coherent two-exciton
recombination. Hence, the induced two-photon emission can lead to the 
appearance
of spectral lines on frequencies $2\Omega-n\omega^s_0-\omega$, where $n$ is
an arbitrary integer number.
\end{references}
\end{document}